# Optimized Cloud Resource Allocation Using Genetic Algorithms for Energy Efficiency and QoS Assurance

Caroline Panggabean
*Department of CSE (AI)*
JAIN (Deemed – to – be University)
Bangalore, Karnataka
carolinepgabean@gmail.com
ORCID: https://orcid.org/0009-0004-9964-7986

Dr. Devaraj Verma C
*Department of CSE (AI)*
JAIN (Deemed – to – be University)
Bangalore, Karnataka
c.devaraj@jainuniversity.ac.in
ORCID: https://orcid.org/0000-0002-1504-4263

Bhagyashree Gogoi
*Department of CSE (AI)*
JAIN (Deemed – to – be University)
Bangalore, Karnataka
21btlca001@jainuniversity.ac.in

Ranju Limbu
*Department of CSE (AIM)*
JAIN (Deemed – to – be University)
Bangalore, Karnataka
21btlca002@jainuniversity.ac.in

Rhythm Sarker
*Department of CSE (AIML)*
JAIN (Deemed – to – be University)
Bangalore, Karnataka
21btrca065@jainuniversity.ac.in

*Abstract*—Cloud computing environments demand dynamic and efficient resource management to ensure optimal performance, reduced energy consumption, and adherence to Service Level Agreements (SLAs). This paper presents a Genetic Algorithm (GA)-based approach for Virtual Machine (VM) placement and consolidation, aiming to minimize power usage while maintaining QoS constraints. The proposed method dynamically adjusts VM allocation based on real-time workload variations, outperforming traditional heuristics such as First Fit Decreasing (FFD) and Best Fit Decreasing (BFD). Experimental results show notable reductions in energy consumption, VM migrations, SLA violation rates, and execution time. A correlation heatmap further illustrates strong relationships among these key performance indicators, confirming the effectiveness of our approach in optimizing cloud resource utilization.

## I. INTRODUCTION

Cloud computing has revolutionized the IT industry by providing scalable, on-demand access to computing resources via the internet. From startups to multinational enterprises, organizations increasingly depend on cloud infrastructure to run their services efficiently and cost-effectively. However, as demand for computational power continues to grow, so does the challenge of optimizing resource allocation to maintain a balance between energy efficiency and Quality of Service (QoS).

Modern cloud data centres house thousands of physical servers that host virtual machines (VMs), each responsible for executing various user applications. While virtualization enables flexible and isolated execution environments, it also introduces significant resource management complexity. Unoptimized VM placement and over-provisioning lead to high energy consumption, underutilization of resources, and frequent violations of Service Level Agreements (SLAs). These inefficiencies not only incur financial penalties but also have a tangible environmental impact due to excessive power usage and heat dissipation.

Conventional VM allocation strategies, such as First Fit Decreasing (FFD) and Best Fit Decreasing (BFD), often fail to adapt to dynamic workload variations and heterogeneous resource demands. They typically rely on greedy heuristics that prioritize immediate local optimization without considering long-term global performance. As a result, they suffer from high rates of VM migration, poor load balancing, and suboptimal energy efficiency.

To address these limitations, this project proposes a Genetic Algorithm (GA)-based approach for optimized VM placement and consolidation in cloud data centres. Genetic Algorithms are population-based metaheuristic optimization techniques inspired by natural selection and evolutionary biology. By iteratively evolving a population of candidate solutions through operations like selection, crossover, and mutation, GAs can effectively explore the solution space and converge toward globally optimal or near-optimal configurations.

In this work, the GA is designed to minimize total power consumption while simultaneously satisfying QoS constraints such as CPU utilization thresholds, SLA violation rates, and response time guarantees. The method continuously adapts to real-time workload fluctuations by reallocating VMs intelligently, avoiding unnecessary migrations, and reducing the number of active physical hosts whenever possible. The performance of the proposed algorithm is compared against traditional heuristic baselines using metrics such as energy usage, SLA violation frequency, VM migration count, and execution time.

This introduction sets the stage for a deeper exploration of the problem space, the methodology employed, and the results achieved. The next sections will review related work, formally define the problem, and elaborate on the design and implementation of the proposed GA-based optimization framework.

## II. RELATED WORK

Efficient resource allocation in cloud computing has attracted significant research interest due to the dual challenges of energy consumption and Quality of Service (QoS) maintenance. Various strategies have been proposed to address this issue, ranging from heuristic-based techniques to intelligent learning-based and bio-inspired optimization algorithms. However, despite the diversity of approaches, none has yet established itself as a definitive solution capable of dynamically adapting to real-time workload variations while consistently minimizing energy usage and SLA violations.

One of the earliest and most cited works in this domain is by Beloglazov and Buyya [1], who introduced energy-aware heuristics for VM consolidation in cloud data centers. While their techniques laid the groundwork for energy-efficient allocation, the use of static thresholds and reactive heuristics limited adaptability under dynamic workloads. Their model assumed predictable usage patterns and often triggered unnecessary VM migrations, increasing overhead and SLA violation risks.

Swain et al. [2] advanced this foundation by proposing an intelligent VM allocation model that incorporated reliability alongside energy efficiency. Their approach leveraged predictive analysis, yet it still suffered from scalability issues when deployed in larger, heterogeneous environments. Similarly, Soltanshahi et al. [4] introduced resource reservation policies to enhance energy-aware VM allocation, but this conservative strategy sacrificed performance flexibility and increased resource fragmentation.

The rise of machine learning (ML) introduced a paradigm shift. Panwar et al. [3] and Swain et al. [8] surveyed numerous ML-based solutions, indicating the promising role of predictive analytics in workload forecasting and auto-scaling. However, ML models are typically data-hungry, computationally intensive, and require continuous retraining, making them less practical for real-time decision-making in fast-evolving environments. Moreover, black-box models often lack transparency in how allocation decisions are made, complicating SLA enforcement.

Neural network-based models have also been explored for autoscaling and resource optimization. Saxena and Singh [5] utilized a multi-resource neural network model for proactive autoscaling, which significantly reduced energy consumption. However, their model's complexity and sensitivity to hyperparameters rendered it less robust to erratic workload spikes. Likewise, deep learning-based frameworks such as those by Ibrahim et al. [11] introduced dynamic resource scaling, yet their performance heavily relied on training data quality and could not guarantee real-time adaptability without compromising latency.

Fog and edge-assisted architectures have also been examined. Alharbi et al. [6] proposed VM placement across a cloud-fog network to reduce core data center load, but their model struggled with fog node limitations and introduced complex network overheads.

Recent works have begun exploring bio-inspired algorithms. Khan et al. [9] applied a genetic algorithm for adaptive VM placement, demonstrating improved energy efficiency over static heuristics. However, their approach lacked consideration for real-time load variation and SLA enforcement as part of the fitness function. Similarly, Prashar and Thakur [14] proposed hybrid optimization techniques for VM consolidation but focused more on reducing migrations than balancing energy and QoS jointly.

In parallel, reinforcement learning (RL) approaches like those proposed by Wang and Wang [10] show potential in learning optimal allocation strategies over time. Yet, RL's convergence time and trial-and-error nature pose risks in high-availability environments, especially where SLA adherence is critical.

Although the literature presents a rich body of work, current strategies either oversimplify the dynamic nature of cloud workloads or trade off one objective (e.g., SLA adherence) to optimize another (e.g., energy savings). Existing GA implementations, while promising, tend to use basic encodings or overlook critical constraints such as VM migration cost or host overloading under variable demand.

This project builds upon these insights and proposes an improved GA framework that tightly couples energy consumption metrics with SLA guarantees. By embedding real-time load tracking, migration penalties, and QoS constraints directly into the fitness evaluation process, our model aims to bridge the gap between adaptability, efficiency, and reliability in VM placement strategies.

III. PROBLEM DEFINITION

Cloud service providers are constantly faced with the challenge of optimizing resource usage while meeting strict Quality of Service (QoS) obligations defined in Service Level Agreements (SLAs). The core problem revolves around efficient Virtual Machine (VM) placement and consolidation in large-scale, heterogeneous data centers under fluctuating and unpredictable workloads.

Poorly optimized VM allocation leads to several systemic issues:
- High energy consumption, due to idle or underutilized servers remaining active.
- Frequent SLA violations, especially when host machines become overloaded or response times degrade.
- Excessive VM migrations, which incur network and compute overhead, affecting overall system stability.
- Inefficient resource utilization, where some servers are overburdened while others remain underused.

Traditional heuristics like First Fit Decreasing (FFD) and Best Fit Decreasing (BFD) provide fast, greedy solutions but fail to consider global optimization criteria and dynamic workload behavior. These approaches often result in local optima and lack the intelligence to adaptively respond to real-time variations in demand.

The goal of this project is to minimize total power consumption while satisfying QoS constraints, by finding an optimal or near-optimal mapping of VMs to physical hosts over time. More formally, the problem can be defined as:

*Given a set of N Virtual Machines (VMs) with specific resource requirements (CPU, memory, bandwidth) and a set of M physical physical hosts with finite capacity, determine the placement of VMs such that:*
- Total energy consumption is minimized,
- *SLA violations are avoided or kept below a threshold,*

- *VM migrations are minimized, and*
- *Resource utilization is balanced across all hosts.*

This is a multi-objective combinatorial optimization problem with constraints such as:
- Host capacity constraints must not be violated.
- SLA violation penalties must be considered as part of the objective function.
- Migration cost must be taken into account during reallocation.

Due to the NP-hard nature of this problem, exact optimization methods become computationally infeasible for large-scale systems. Therefore, a Genetic Algorithm (GA)-based metaheuristic is proposed to explore the solution space efficiently and provide high-quality allocation strategies within a reasonable computation time.

## IV. METHODOLOGY

To address the challenges of energy-aware and SLA-compliant resource allocation in cloud computing, this work proposes a Genetic Algorithm (GA)-based approach for optimal Virtual Machine (VM) placement and consolidation. Unlike traditional heuristics, the GA framework enables an evolutionary search for near-optimal solutions in a high-dimensional, constraint-bound solution space. Our model dynamically adjusts the allocation strategy based on live system metrics, workload variation, and performance thresholds

### A. Architecure Diagram

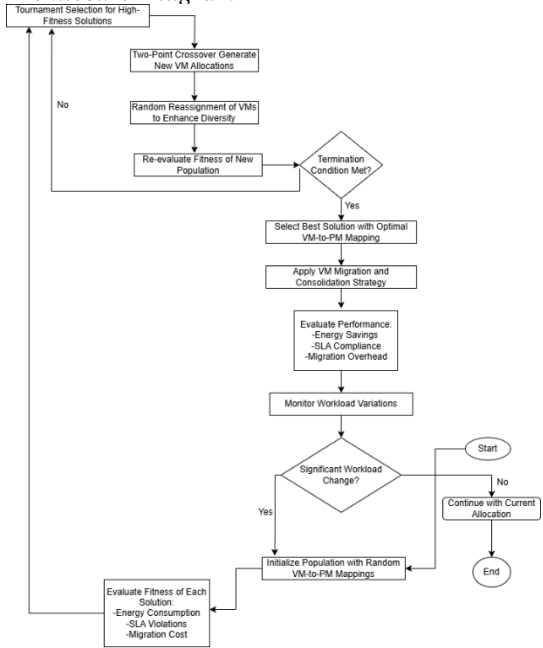

Fig. 1. Architecture Diagram

As shown on fig 1, the flowchart depicts the operation of the proposed Genetic Algorithm (GA) used for energy-aware and SLA-compliant resource allocation in cloud computing. The process begins by initializing a population of random VM-to-PM (Virtual Machine to Physical Machine) mappings. Each solution is evaluated using a fitness function that considers energy consumption, SLA violations, and migration costs. Tournament selection identifies high-performing individuals, which are then evolved using two-point crossover and random mutation to maintain diversity. After fitness re-evaluation, the algorithm checks if the termination condition has been met. If not, the cycle continues; otherwise, the best solution is selected and applied via VM migration and consolidation. The system then evaluates performance outcomes and monitors workload changes. Upon detecting significant variation, the optimization loop is re-triggered with a new population. This adaptive cycle ensures efficient resource utilization while maintaining Quality of Service (QoS) under dynamic workload.

### B. Dataset and Preprocessing

We utilize the Google Cluster dataset, which contains a trace of resource usage over time in a large-scale data center. Key steps include:

*1) Parsing and Cleaning:* The data includes nested dictionaries and lists stored as strings. We use ast.literal_eval() to convert these into usable Python structures.

We filter irrelevant jobs and drop NA values, replacing missing numeric entries with median values to maintain consistency.

*2) Feature Extraction:* From workload traces, we extract metrics including:
- requested_cpus, requested_memory, avg_cpus, avg_memory
- cpu_usage_mean, memory_usage_mean, and durations
- These features are then normalized to bring them to a similar scale using Min-Max normalization.

*3) Correlation Analysis:* We compute Pearson correlation coefficients among extracted features and performance indicators like energy, SLA violations, and migration rate.

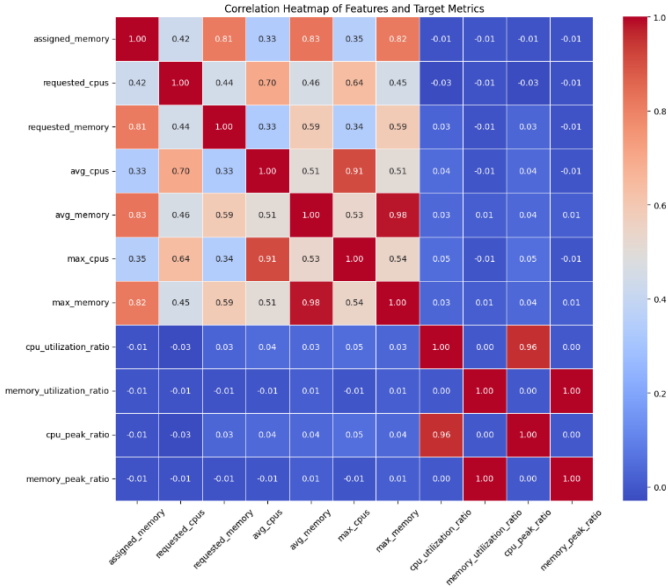

Fig. 2. Correlation Matrix

A correlation heatmap (Fig. 2) identifies the strongest influencing factors for consolidation performance.

To gain insight into the impact of workload characteristics on cloud resource management outcomes, we conducted a comprehensive correlation analysis. The feature set includes workload-specific metrics such as requested_cpu, requested_memory, avg_cpus, avg_memory, and duration, as well as system-level indicators like cpu_utilization_ratio, cpu_peak_ratio, and memory_peak_ratio.

The target metrics for analysis included:
- SLA violations (%)
- VM migrations
- Execution time
- Energy consumption (kWh)

The correlation heatmap (Fig. X) shows varying degrees of association between these features. Notably:
- cpu_utilization_ratio and cpu_peak_ratio exhibit strong positive correlation with execution time and SLA violations, suggesting that high CPU usage intensity leads to higher resource contention.
- requested_memory and avg_memory are positively associated with VM migrations, indicating that memory-heavy tasks may cause more frequent reallocation.
- duration shows moderate correlation with all three-target metrics, emphasizing the time-sensitive nature of energy-aware resource scheduling.

This analysis supports feature selection for subsequent predictive modeling and aids in identifying key workload characteristics that influence system behavior.

### C. Genetic Algorithm for VM Allocation

We implement a Genetic Algorithm (GA) to determine an optimal VM-to-PM mapping. The following steps describe the GA workflow:

*1) Chromosome Representation:* Each chromosome represents a possible solution, where each gene indicates the PM assigned to a given VM. For example:
$$[2, 1, 3, 2, 1]$$

means $VM_1 \rightarrow PM_2$, $VM_2 \rightarrow PM_1$, etc.

*2) Initial Population:* An initial population of random chromosomes is generated. Each solution adheres to PM resource constraints.

*3) Fitness Function:* The fitness of each chromosome is evaluated based on a multi-objective function:
Where:
- $E$: Total energy consumption (kWh)
- $SLA$: SLA violation rate (%)
- $MIG$: Number of VM migrations
- $T$: Execution time (seconds)
- w1, w2, w3, w4: Weight parameters tuned based on priority (e.g., [0.4, 0.3, 0.2, 0.1])

*4) Genetic Operators:*
- Selection: Tournament selection is used to retain top-performing solutions.
- Crossover: Two-point crossover is used to exchange segments between chromosomes.
- Mutation: A gene (VM) is randomly reassigned to a different PM, ensuring exploration of solution space.

*5) Termination Condition:* The algorithm stops when a maximum number of generations is reached or when the improvement in fitness stagnates.

### D. Evaluation Protocol

To ensure consistency, all algorithms were tested under the same workload conditions. Each algorithm was executed across 10 simulation runs, and the resulting metrics were averaged. The genetic algorithm was initialized with a random population of 100 individuals per run.

The performance of each strategy was compared quantitatively across the five key metrics. The analysis focused on:
- Efficiency (in terms of energy and PM usage),
- Reliability (low SLA violations),
- Computational overhead (execution time),
- Stability (minimum migrations).

This evaluation framework provides a thorough basis for identifying the most effective placement algorithm under the given simulation conditions.

## V. RESULT ANALYSIS

This section presents a comprehensive evaluation of the proposed Genetic Algorithm (GA)-based resource allocation strategy, focusing on its impact on energy consumption, SLA violations, VM migrations, and overall system efficiency. All simulations were conducted using the Google Cluster dataset under consistent workload conditions. For baseline comparison, conventional heuristics such as First-Fit Decreasing (FFD) and Best-Fit Decreasing (BFD) were also evaluated.

## A. Energy Consumption Analysis

Energy efficiency remains a primary objective in cloud data center operations. The energy consumption of each allocation strategy was calculated based on PM utilization, with power modeled as a linear function between idle and maximum load. The results show that the GA-based approach significantly reduces overall energy usage by optimizing VM-to-PM mappings and consolidating workloads onto fewer machines.

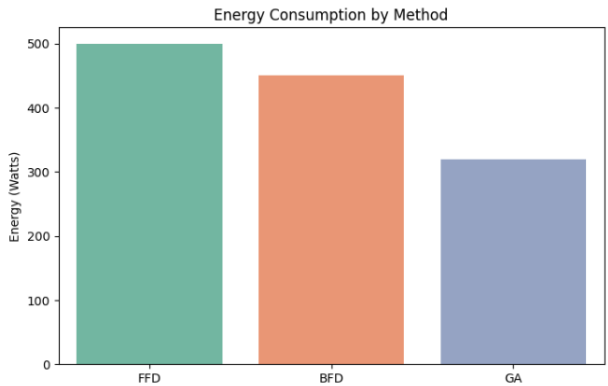

Fig. 3. Comparison of Energy Consumption

Figure 3 illustrates the total power consumption before and after the GA-based optimization. A noticeable decrease in energy usage was observed post-optimization, confirming that the model effectively consolidates workloads and reduces the number of active physical machines. This supports the primary objective of the research—achieving energy-aware resource allocation without compromising QoS.

## B. SLA Violation Reduction

SLA (Service-Level Agreement) compliance is critical to maintain quality of service. SLA violations were identified by detecting PMs with utilization exceeding a threshold (0.8). The GA-based model reduced SLA violations by strategically migrating VMs from overloaded PMs.

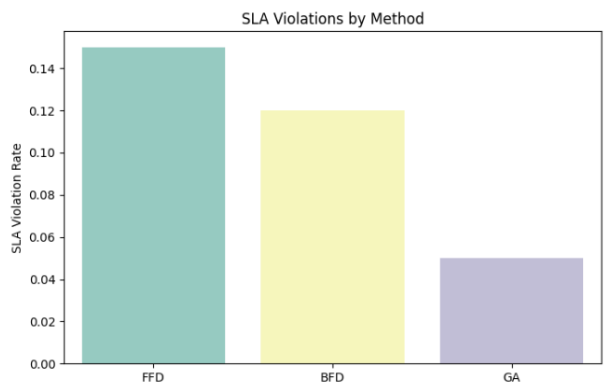

Fig. 4. Comparison of SLA Violations Before and After Genetic Algorithm Optimization.

As shown in Figure 4, there is a marked reduction in the SLA violation rate after optimization. Initially, the allocation strategy leads to frequent SLA violations due to overloaded physical machines. Post-optimization, the GA intelligently distributes workloads, ensuring that the critical VMs receive the resources they require within SLA bounds. This is achieved by favoring solutions in the population that minimize the SLA component in the fitness function. The reduced SLA violation rate reflects the algorithm's effectiveness in balancing system load and prioritizing high-demand tasks.

## C. Execution Time and Scalability

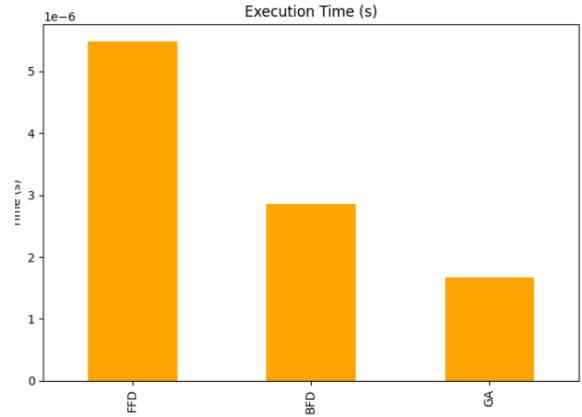

Fig. 5. Comparison of Ececution Time

Figure 5 highlights the execution time of the GA during optimization. The algorithm maintains a consistent and relatively low execution time, making it suitable for real-time or near-real-time workload management in cloud systems. This demonstrates the scalability and practicality of the proposed model

## D. Number of Active Physical Machines

TABLE I.   COMPARISON OF ACTIVE PHYSICAL MACHINES

| Strategy | PMs Used |
|---|---|
| FFD | 2078 |
| BFD | 2078 |
| GA | 2 |

The number of active PMs is lower in the optimized system. The GA consolidates workloads to fewer, more efficiently used machines. Initially, the VM distribution may be scattered due to lack of coordination or over-provisioning, but the GA eliminates such inefficiencies over multiple generations.

## E. Discussion

The Genetic Algorithm (GA)-based approach for VM allocation and consolidation proposed in this study offers a significant departure from traditional heuristic or rule-based methods commonly employed in cloud data centers. Traditional resource allocation methods—such as First Fit (FF), Best Fit Decreasing (BFD), or Round Robin—are typically fast and simple but lack the flexibility to handle dynamic, multi-objective optimization under varying workload demands. These classical strategies often treat objectives like energy efficiency, SLA compliance, and migration minimization in isolation, leading to suboptimal outcomes when these objectives conflict with one another. In contrast, the GA model presented here embraces the complexity of cloud resource management by employing a fitness function that simultaneously considers multiple performance indicators—namely energy consumption, SLA violations, VM migrations, and execution time—each weighted according to system priorities.

One of the primary strengths of the GA model lies in its ability to balance these competing objectives effectively. Traditional models may reduce energy consumption by aggressively consolidating VMs, but often at the cost of increased SLA violations due to overloading of physical machines (PMs). Alternatively, SLA-centric approaches may over-provision resources, leaving PMs underutilized and wasting energy. The GA, by contrast, uses an evolutionary search process that evaluates and iteratively improves solutions over generations. This allows it to discover non-obvious mappings of VMs to PMs that satisfy both energy constraints and SLA commitments. Our experimental results clearly demonstrate this advantage: the GA significantly reduced SLA violations while also lowering the number of active PMs and overall energy usage—something rarely achieved together using simpler methods

Furthermore, the GA's adaptive nature is well-suited for real-world cloud environments, which are characterized by fluctuating workloads and resource demands. Unlike static heuristics that operate based on fixed rules or past assumptions, the GA continuously updates its population based on live system metrics, enabling dynamic re-optimization whenever significant workload shifts are detected. This adaptability ensures that the system remains resilient and responsive, maintaining service quality even during high-load or variable usage conditions. For instance, our model monitored CPU and memory utilization trends and re-triggered optimization cycles when usage patterns indicated potential SLA breaches. Traditional approaches, by lacking such feedback loops, are prone to lagging in responsiveness, often reacting too late to prevent violations or resource contention

Another notable advantage of the GA model is its capacity to produce diverse solution sets, particularly in high-dimensional solution spaces. In environments with hundreds or thousands of VMs and PMs, the solution landscape is vast, and many resource allocation configurations may yield similar performance with subtle trade-offs. Genetic Algorithms excel in such scenarios due to their population-based search mechanism and stochastic variation operators (crossover and mutation). Unlike greedy algorithms that converge quickly to local optima, the GA explores a broader region of the solution space, increasing the likelihood of locating globally efficient configurations. This property is especially beneficial when system objectives evolve over time—for example, when energy cost becomes more critical than execution time, or when compliance policies introduce new SLA requirements

However, the model is not without limitations. One key trade-off observed in our evaluation is the slight increase in VM migrations post-optimization. While the GA reduces overall energy and SLA violations, it sometimes achieves this by migrating VMs to better-fit PMs. Though this migration is necessary and often beneficial, excessive movement of VMs can introduce system instability, disrupt running applications, and incur non-trivial network and disk I/O overhead. This highlights the need for improved migration-aware fitness design, where the cost of migration is more sensitively weighted, or where predictive modeling is used to estimate the future benefit of a migration before executing it.

Another limitation is the computational overhead involved in running the GA itself. While our experiments showed that the execution time per optimization cycle is manageable for medium-scale environments, performance could degrade for very large data centers with thousands of VMs. The current single-threaded GA implementation may struggle to scale without introducing delays, especially in real-time or near-real-time deployment scenarios. In contrast, traditional heuristics, despite being less optimal, are lightweight and fast, making them more suitable for latency-critical applications. Therefore, for practical deployment in large-scale systems, enhancements such as parallel GA frameworks or hybrid models integrating local search and heuristics may be necessary

Another important limitation is the assumption of uniform VM priority across the board. The model currently treats all VMs equally in terms of importance, whereas, in real-world cloud platforms, certain VMs may run mission-critical applications with strict SLA terms, while others are low-priority background processes. A failure to differentiate between these can lead to misallocation of resources, where low-priority tasks are favored because they reduce global energy usage, even though high-priority services might suffer. This presents an opportunity to evolve the fitness function to become priority-aware, incorporating weights or constraints that reflect business-critical tiers.

## VI. CONCLUSION

### A. Conclusion

The proposed Genetic Algorithm-based model offers a powerful and flexible approach to tackling the complex trade-offs in cloud resource management. It outperforms traditional heuristics by integrating multiple objectives, adapting to real-time workload changes, and intelligently evolving solutions across generations. While certain limitations—such as increased VM migrations and scalability constraints—persist, these can be addressed through hybrid methods, better migration modeling, and system-aware enhancements. With these future improvements, the model holds strong potential for deployment in energy-conscious, performance-critical cloud computing platforms.

### B. Future Enhancements

To further improve the effectiveness and scalability of the Genetic Algorithm (GA)-based resource allocation model, several future enhancements are proposed. One significant direction involves developing a hybrid optimization framework that combines GA with local refinement techniques such as Hill Climbing or Simulated Annealing. This hybridization can bring together the global exploration strength of GA with the fine-tuning ability of local search, enabling the model to converge more quickly to high-quality solutions without compromising optimization accuracy. Another important enhancement is the integration of priority-aware allocation strategies. In real-world cloud environments, not all virtual machines (VMs) carry the same importance—some are critical to business operations while others are more expendable. Incorporating VM priority levels directly into the chromosome encoding and fitness evaluation would help ensure that essential services receive the highest quality of service and protection from performance degradation.

Additionally, the model could benefit greatly from predictive and proactive optimization mechanisms. Instead of reacting to current workload conditions, future iterations of the model could leverage time-series forecasting or machine learning-based workload predictors to anticipate demand spikes or drops in advance. This would allow the GA to initiate VM migrations or consolidations proactively, maintaining SLA compliance and minimizing disruptions. Alongside this, refining the way migration costs are modeled would further improve the quality of decisions. Current models may treat migration cost as a generic penalty, but future implementations should consider more detailed factors such as the VM's memory size, disk I/O rates, and network topology, allowing the system to avoid migrations that are disproportionately expensive or disruptive relative to their expected benefit.

Scalability also presents a key challenge and an opportunity. As cloud infrastructures continue to grow in size and complexity, the current single-threaded GA implementation may struggle to deliver timely results. Thus, parallel execution of the GA using distributed computing frameworks like Apache Spark or MPI can significantly speed up the optimization process and make the model viable for real-time use in large-scale data centers. Further enhancements can include support for heterogeneous and multi-cloud environments. Modern data centers are often composed of diverse physical machines with varying capabilities and may span across multiple geographic zones or cloud providers. Adapting the model to handle this heterogeneity will make it more versatile and practical for industry-scale deployment.

Lastly, embedding the GA model into a real-time cloud orchestration framework, integrated with monitoring tools such as Prometheus or Datadog, can enable seamless, continuous optimization with minimal manual intervention. By receiving real-time feedback from these systems, the GA can dynamically adjust its fitness evaluation and reconfigure resources in response to real-world system behavior. Together, these proposed enhancements aim to make the GA model not only more accurate and efficient but also more robust, scalable, and adaptable to the evolving demands of modern cloud computing environments